# Is there a Darwinian Evolution of the Cosmos?

## Some Comments on Lee Smolin's Theory of the Origin of Universes by Means of Natural Selection

Rüdiger Vaas

*Zentrum für Philosophie und Grundlagen der Wissenschaft*
*Justus-Liebig-Universität Gießen, Germany*
*Ruediger.Vaas@t-online.de*

> „Many worlds might have been botched and bungled, throughout an eternity, 'ere this system was struck out. Much labour lost: Many fruitless trials made: And a slow, but continued improvement carried out during infinite ages in the art of world-making."
>
> David Hume (1779)

## 1 Abstract and Introduction

For Lee Smolin [54-58], our universe is only one in a much larger cosmos (the Multiverse) – a member of a growing community of universes, each one being born in a bounce following the formation of a black hole. In the course of this, the values of the free parameters of the physical laws are reprocessed and slightly changed. This leads to an evolutionary picture of the Multiverse, where universes with more black holes have more descendants. Smolin concludes, that due to this kind of Cosmological Natural Selection our own universe is the way it is. The hospitality for life of our universe is seen as an offshot of this self-organized process.
This paper outlines Smolin's hypothesis, its strength, weakness and limits, and comments on the hypothesis from different points of view: physics, biology, philosophy of science, philosophy of nature, and metaphysics.

---

# 2      Explananda and conjectures

## 2.1     Crucial questions

• Why do the about 20 free parameters of the standard model of elementary particle physics – and also some important cosmological parameters (expansion rate, deceleration parameter, mean mass density, cosmological constant, inflation charge, kind and percentage of nonbaryonic matter etc.) – have the values they actually have?
• Why are the parameters „fine-tuned" for (earth-like) life?

## 2.2     Some proposals for an answer

The universe is the way it is
• due to a result of pure chance, and there is no possibility and no need for an explanation at all,
• due to a result of pure chance, because every set of parameters is realized somewhere in a cosmos (Multiverse) of many universes; we happen to live in a universe which must be hospitable for (earth-like) life due to the parameters, which are a necessary conditions for such forms of life, because otherwise we could not observe these physical properties and ask this question (Weak Anthropic Principle),
• because otherwise we would not exist (Strong Anthropic Principle),
• because there is a final Theory of Everything which necessitates exactly these values of the parameters, i.e. there are, as a matter of fact or logic, no other values possible – hence, there are no „free" parameters at all,
• because our universe was designed to have these very values (Cosmological Argument from Design),
• because the values of the parameters were a product of a Cosmological Natural Selection, i.e. the result of a self-organized process.
The last kind of answer is Smolin's account.

## 2.3     Universes within the Multiverse

There are different meanings of the term „universe" [33, p. 277], for example:
• Everything (physically) in Existence, ever, anywhere,
• the region we inhabit plus everything that has interacted or ever will interact with this region,
• this region plus everything that has interacted with it by now, or will at least do so in the next few billion years,
• any gigantic system of causally interacting things that is wholly (or very to a very large extent) isolated from others,

• any system that might well have become gigantic, etc., even if it does in fact recollapse while it is still very small.

Nowadays, the term „cosmos" or „Multiverse" might be used to refer to Everything in Existence, while „universe" is used (and will be used here) in a way permitting talk of several universes within the Multiverse. In principle, these universes might or might not be spatially, temporally and/or dimensionally separated from each other.

## 2.4 Cosmological Natural Selection

### 2.4.1 The argument

Smolin's hypothesis depends on the following argument:

*If* a new universe is born from the center of a black hole, i.e. if the bounce of a singularity leads to a new expanding region of spacetime,

*and if* the values of the fundamental free parameters of physical theories can change thereby in a small and random way, i.e. differ from those in the region in which the black hole formed (in particular, Smolin has in mind the dimensionless parameters of the standard model of particle physics),

*then* this results in different reproduction rates of the physically different universes.

*Hence*, our universe has been selected for a maximum number of black holes. It is a descendant of a series of universes, each of which had itself been selected for the same criterion.

*Thus*, the values of the parameters are the way they actually are, because this set of values leads to a (local) maximum of descendant universes.

### 2.4.2 Premises

This line of reasoning depends on the following premises:

• The gravitational collapse of matter (and, especially, of the core of an exploding star) does not end in a deadlock singularity inside a black hole but gives birth to an offspring universe.

• While the form of the laws of physics remains the same, there are small and random changes of the values of the laws' parameters during the bounce which leads to the origin of a new universe with slightly different fundamental properties.

These premises of an evolution of universes and changes of physical constants and laws are not entirely new [cf. e.g. 35; 47, ch. 44; 76, p. 239 ff]. The possibility of a bounce, although speculative, has been even discussed recently in several different approaches to quantum gravity, including string theory [20; 31; 42]. The novelty of Smolin's approach is the introduction of *small* changes of the parameters' values and his application of Darwinism to cosmology. (Radical changes are not very interesting because in those universes which differ substantially from our own, stars like ours could not form – the conditions for star formation are rather special).

### 2.4.3 Auxiliary assumption

There is also an important auxiliary assumption:
• Our universe is typical, i.e. nearly optimal for black hole formation.

### 2.4.4 Implications: the evolution of universes

Now, the implications are:
• Given the premises, it is probable that a universe chosen at random of a given collection of physically possible universes has parameters that are near a peak (local extremum) of the formation of black holes after a sufficient time (or, if we can't or won't define a multiversal i.e. trans-universal time: after a sufficient number of generations).
• Given also the auxiliary assumption, i.e. if our universe is a typical member of that collection of physically possible universes, then its fundamental parameters must be close to one of the peaks of the formation of black holes.
• Thus, most changes in the values of the parameters of the physical laws would *decrease* the rate at which black holes are produced in our universe. That is, most universes with (even only slightly) different constants of nature compared to our own universe would contain less black holes. – *This conclusion is testable according to Smolin. And he actually suggested some ways how to test it (see below).*
• Thus, Smolin's answer to the question regarding the values of the physical parameters is simply this: The parameters have the values we observe, because these make the formation of black holes much more likely than most other values.
• Hence, according to Smolin our universe is a product of mutation and selection analogous to the evolution of species described first by Charles Darwin in his seminal book „On the Origin of Species" in 1859 [12] and independently also by Alfred Russel Wallace [cf. 74].

# 3 Science

## 3.1 Physics and cosmology

### 3.1.1 Which parameters?

Of course it is not known if we are varying the right parameters to test Smolin's hypothesis. It is unclear, how many free parameters there are, whether their variation is random etc.
But these open questions are no objections. As long as there is no better tool at hand, everything we have should be tried. And with increasing knowledge about the fundamental parameters the hypothesis of Cosmological Natural Selection could be improved.

### 3.1.2 Refutation by a Theory of Everything?

Smolin's hypothesis is not inconsistent with the hypothesis that there is a fundamental physical theory. Thus, the possibility of the discovery of such a Theory of Everything is no objection but would be another and perhaps an even better opportunity to test his hypothesis.

However, Smolin's hypothesis is inconsistent with the idea that a fundamental theory could *uniquely* determine the observed parameters of the standard model of particle physics and cosmology. If it could be shown that the values of the parameters must be the way they are as a matter of fact or logic, there still could be a Multiverse consisting of indefinitely many universes (with the same laws and parameters like ours) but no natural selection of these universes, hence no cosmological Darwinian evolution.

### 3.1.3 Open questions regarding new universes born from black holes

Even if black holes are places of birth for universes it is not clear
• whether the values of the physical parameters really varies by *small* amounts and *randomly* as it is presumed,
• what happens to the already born universes if their mother black holes merge together or evaporate due to quantum mechanical events (Hawking radiation),
• whether there are further universes created if black holes merge together – and how many: one or two?
• why there is only *one* offspring from a black hole and not (infinitely) many,
• and, if the latter is true, whether the numbers of new universes which are born from each black hole may differ according to the mass of the black hole. (Claude Barrabès and Valerie Frolov [2] have suggested that a large number of universes might be created inside each black hole and that the number of universes produced that way may grow as the mass of the black hole increases. If so, universes should be selected for *supermassive* black holes, not for sheer numbers of black holes.)

It is also not clear whether the different universes interact which each other [cf. 8; 9]. There is even the threat of a reciprocal destruction.

Another restriction of Smolin's approach is that his cosmic reprocessing mechanism only leads to different values of parameters, but not to different *laws*. His hypothesis still requires the same basic structure of the laws in all the universes in order to make sense. But of course an even more radical proposal – a variation not only of universal constants but also of universal laws – is beyond every possibility of scientific investigation (at least for now). Furthermore, we simply do not know if and where a distinction is useful between universes which are physically possible, as opposed to those that can only be imagined (which are only conceivable or logically possible).

### 3.1.4 Predictability and Testability

See below (5.3).

## 3.2 The status of life

### 3.2.1 There is no necessary connection between black holes and life

In principle, life and Cosmological Natural Selection could be independent of each other. There are two reasons for this:
• On the one hand there may be universes full of black holes were life as we know it couldn't evolve. For example it might be possible that there are only short-lived giant stars which collapse quickly into black holes, or that there are universes dominated either by helium or by neutrons (corresponding to the neutron/proton mass difference being either zero or negative), or that there are universes with many (more) primordial black holes [27; 36; 78-80] and maybe without stars at all. Such universes might be very reproductive because of their giant stars or primordial black holes but are not able to produce earth-like life.
• On the other hand we can conceive a universe without black holes at all (if supernovae lead to neutron stars only, or if there are not stars above a critical mass limit) but which could be rich in earth-like life nevertheless.
Thus, there is not a (logically) necessary connection between black holes and life [cf. 18; 49].
Smolin's hypothesis of Cosmological Natural Selection therefore cannot necessarily explain the presumed fine-tuning of the universe for life. By the way, Smolin [57, p. 393] also stressed that both properties of our universe – containing life and producing a maximum number of black holes – must be taken as independent for the purpose of testing the theory.

### 3.2.2 Nevertheless there may be a contingent connection

Namely via the role of carbon [57; 58]
• as the „molecule of life", because of its ability to make complex molecules (to a much larger extent than any other element),
• as an element accelerating star formation, because of the role that carbon monoxide plays in shielding and cooling the giant molecular clouds of gas and dust were stars are born.
Thus, there may be at least a coincidence between the conditions for maximizing black hole formation and being hospitable for life. If Smolin is right, then this could even be a physically necessary connection. But we must still wonder why the laws of nature are such that this linkage occurs. (And it is not clear, whether carbon monoxide really does increase the number of black holes, because hydrogen cools efficiently too, and all the stars in the early universe were carbon-free [52].)

### 3.2.3 Anthropic considerations

According to the so-called Anthropic Principle [see 1; 3; 4] the following is true: „what we can expect to observe must be restricted by the conditions necessary for our presence as observers" [7, p. 291] resp. „We see the universe the way it is because we exist" resp. „... because if it were different, we would not be here to observe it" [27, p. 124, 183].

To give a *causal explanation* for the fine-tuned parameters (given that life is actually restricted to some special conditions [cf. 62]),

• there must be either a final theory to necessitate this fine-tuning (or, if one wishes to avoid a Platonic interpretation, the fundamental properties of nature are as a matter of fact in accordance with a quite high probability of the existence of life, i.e. the view that the properties of our universe are enormously improbable is mistaken – hence, strictly speaking there is no fine-tuning at all),

• or there must be a selection effect at work, which could consist in an Observational Selection due to the existence of many different universes, or it could consist even in a Cosmological Natural Selection of universes which are hospitable for life.

Thus, if it could be shown that a maximum of black hole formation rate is indeed correlated with (or followed by) a hospitality for life, Smolin's hypothesis would indeed causally explain the fine-tuning. (Note that Cosmological Natural Selection implies Observational Selection but not vice versa.)

### 3.2.4   Life and Intelligence as an epiphenomenon

If they do not contribute to black hole formation, life and intelligence are a mere byproduct or epiphenomenon in Smolin's scenario, i.e. they are causally inert (regarding the evolution of the universes). Thus, our universe was not positively selected for life, even if the conditions of life would be exactly the same as the conditions for maximizing the numbers of black holes. Therefore, Cosmological Natural Selection does not imply (or entail) a „meaning" or function or advantage of life.

### 3.2.5   Science-fiction: making other universes

But couldn't it be possible, that, nonetheless, there is a hidden connection between the hospitality of universes for life on the one hand and black hole formation on the other? Perhaps black holes could be advantageous for life, or life could be advantageous for black hole formation.

For instance, successful industrial civilizations will eventually create black holes as waste disposals, weapons, accelerators and research tools [68]. Or they need small and therefore hot black holes for energy production and as heaters to keep warm and alive in a forever expanding, decaying universe which runs out of free energy [11]. In principle, black holes could provide a permanent solution to this problem, since they convert matter into energy with perfect efficiency via the Hawking radiation. Thus, it is possible that life creates other universes as an unintentional by-product of advanced technologies. Furthermore, it was speculated that intelligent cosmic engineers could create universes by means of black holes [19; 20; 24]. Maybe they use the other universes as laboratories or as weird zoos. Or they need them as future homes, if their own universe starts to collapse or is burned out [13, pp. 149 f]. In that case, they also need traversable wormholes to change places [70; 73]. There would be no self-organized evolution in this case but rather a preplanned development. Nevertheless, life could still be seen as a „tool" of the Multiverse to produce more universes. However, it would be no epiphenomenon in this case but like „medusids, and black holes/universes are the polyps of a single evolutionary process" [11, p. 5].

Of course these suggestions are pure, speculative science-fiction at the moment – in line with Olaf Stapledon's novel „Star Maker" [59]. Nevertheless, the speculations could be verifiable in principle – if, for instance, a hidden message could be deciphered in the exact values of fundamental parameters or numbers like Pi [50, ch. 24] – a greeting card from some strange cosmic engineers who created our universe in their laboratory.

# 4 Terminology

## 4.1 The Darwinian analogy is an inadequate model transfer

Smolin's Darwinian analogy of Cosmological Natural Selection is in some respect misleading. It is useful to define Natural Selection with Smolin as the selective growth of a population on a fitness landscape with the landscape's altitude determining the rate of reproduction. But this is not sufficient, because natural selection as described in biology – there is a broad consensus among biologists here, [cf. e.g. 15, ch. 2; 16, ch. IV; 21, ch. 6, 7; 22, part III; 25; 34, ch. 2; 37; 44, part II; 45, p. 97; 46, ch. 11; 48, ch. 4; 77, ch. 7] – also depends on the assumption that the spread of populations (or genes) is crucially constrained by *external* factors (shortage of food, living space, mating opportunities etc.). The fitness landscape is shaped not only by internal restrictions, but also by external ones – there would be no adaptation without them. Resources are limited – this is a crucial premise of Darwin's theory (already stressed by Thomas R. Malthus [38] who had an important influence on Darwin's thinking, and, independently, also on Alfred Russel Wallace, the cofounder of Darwinism, [cf. 74]). In comparison with that, the fitness of Smolin's universes is constrained by only one factor – the numbers of black holes –, and this is an *internal* limitation. Although Smolin's universes have different reproduction rates, they are *not competing against each other* [43]. Although there are „successful" (productive) and „less successful" universes, there is no „overpopulation" and no selective pressure or „struggle for life" resp. „struggle for existence" as Darwin has put it [12, p. 459, 490] (somewhat misleading, by the way, but he meant the term „in a large and metaphorical sense"). Hence, there is no natural selection in a strict biological (Darwinian) sense. Smolin's universes are isolated from each other (except maybe for their umbilical cords). Therefore, there couldn't be a quasi-biological evolution of universes.

However, Smolin may still insist that competition has nothing essential to do with natural selection and could give arms races and symbioses of species as evolution's counter-examples. But these counter-examples are, contrary to

Cosmological Natural Selection *and* a broad consensus among biologists, also based on a (partially shared) common environment which is continuously interacting with the species in question. And ultimately, arms races as well as symbioses are also based on competition. For instance, arms races are struggles for resources, e.g. food or shelter, dominance or mating partners; and symbiosis – which can also be seen as reciprocal parasitism – is competition between the symbionts against other individuals of their species with less effective symbionts, or competition between different species which try to get the same resources, or even competition for protection mechanisms. By the way, there is nothing like an arms race or symbiosis between different universes. Thus, biological natural selection in general and Darwinian natural selection in particular crucially depends on external interactions and competition – and this is not just a matter of definition.

In conclusion: „Cosmological Natural Selection" is a somewhat misleading term and an inadequate model transfer. But of course this does not prove Smolin's hypothesis wrong.

## 4.2 There is no life of the cosmos

It is even more problematic to compare or equalize universes (or the Multiverse) with life [23] – an idea, by the way, that goes back at least to Plato (4th century BC), cf. his „Timaios" 92 b.

If we take life as a self-organized, nonequilibrium system governed by a symbolic program (like the genetic code [64]), which can reproduce both itself and its program, as Smolin does [57, p. 194], Smolin's universes may fit into this definition of life, although he does not „see that it is useful" [personal communication] and does not want to stretch the analogy to this extreme [57, p. 195]. But if one applies his definition of life to our universe, one might indeed be forced to conceptualize it as alive. For, if Smolin is right, self-reproduction and some sort of a self-organized non-equilibrium seem to be properties of our universe (expanding or contracting universes are not in equilibrium, and they are also not perfectly closed systems due to their relationships with other universes). If we take the parameters and laws of nature as a program which is stored symbolically (i.e. which is realized or represented by physical events), all three premises of Smolin's definition of life cited above are satisfied for the universe.

However, Smolin does not want to regard the laws of nature as a program because „there is no sense in which the laws of nature could be represented symbolically, as something analogous to a computer program" [57, p. 195]. But this is not the point. Biological life does not care if computer programs or whatever can represent the genetic code symbolically. It is even misleading to treat the genetic code as real (as objective information *sui generis*) – it is *our* abstract description and, in fact, simplification of the complex causal interrelationships of the biochemical processes within a cell which we call transcription and translation [64, p. 44 f]. Likewise, we need not conceptualize nature's parameters and laws as a program in the strict sense. If we do not want to give them a Platonic or a purely instrumentalistic (anti-realistic) status, we can think of them as abstract descriptions, as our hypotheses of observed regularities which are a result of specific mind-independent properties of nature – a view which seems to be compatible with Smolin's relational picture of the cosmos.

To avoid the view that the universe is alive, Smolin's definition of life has to be extended. And there is something missing indeed: *interaction with an environment*. Life as we know it is in a permanent exchange with its surroundings. This is not true for Smolin's universes. Here, there are only single one-way effects (births), nothing more – no interactions in either way. This difference is already sufficient not to think of our universe as a living system.

Furthermore, life, as we know it, depends both on living and nonliving entities to interact with. This is also not true for Smolin's universes. Here by his own definition, only (living or non-living?) universes are surrounding our universe. Hence, the metaphor of a living cosmos is a mixture of opposite terms.

Finally, if it is already strange to conceptualize living universes, why should we even conceive the whole cosmos (Multiverse) as a living one? It is by definition not in an exchange with an environment, and it is by definition not an offspring of reproduction.

In conclusion: Neither our universe nor the Multiverse is alive if „alive" is understood in terms of ordinary language and biology.

But of course this terminological criticism is no objection to Smolin's basic ideas.

# 5 Philosophy of Science and Nature

## 5.1 Advantages and Consequences

### 5.1.1 A truly Perfect Cosmological Principle and the ultimate expulsion

Suppose, Smolin is right. The implications for our view of nature would be enormous. Our universe would be only a grain of sand on the incredible large beach of the Multiverse. And it would be in no way special.

According to the *Cosmological Principle*, our universe is homogeneous and isotropic on the large scale, that is, it appears the same at all places and, from any one place, looks the same in all directions. If Smolin is right, we could accept a truly *Perfect Cosmological Principle*, that is, the Multiverse is literally „full" of universes like our own universe.

It has often been said that the Copernican revolution has catapulted the earth and hence, mankind, out of the center of the universe. In the twentieth century it became clear that neither the earth, nor the sun, nor the milky way, nor the local supercluster of galaxies are at the center of the universe, because there is no center at all. Nor does the baryonic matter of which we consist dominate the universe, i.e. in comparison with the assumed particles of dark matter even the material we are made of is quantitatively negligible. This dramatic widening of our horizon and diminution of our role in the universe led to a radical cosmic expulsion. Our posi-tion in space-time is completely irrelevant, providing no evidence for a universal meaning or a cosmic value of mankind, nor a protection against contingency and absurdity [67]. In recent decades, some interpretations of the Anthropic Principle and the alleged fine-tuning of the physical parameters have been examples for a tendency to reverse this development. But the idea of the Multiverse made out of many different universes, which implies an Observational Selection effect, is once again sufficient to wipe out any romantic dreams of anthropocentrism. But in these picture at least our life-bearing universe might still be special. In Smolin's account however, our universe has most properties in common with all other universes. It would be a very ordinary world indeed. Thus, if Smolin is right, there is no reason to believe that (human) life is unique or that the constants of nature have a singular status.

### 5.1.2 Advantages of Smolin's hypothesis

Given the fact that there is no fundamental theory which necessitates the values of the fundamental parameters, Smolin's hypothesis is the only one around which is able to give a causal explanation of them – in opposition to non-scientific teleological speculations on the one hand and the claim of pure accident or a lucky chance on the other, which are not falsifiable resp. verifiable in principle and hence not scientific at all.
Furthermore, Smolin's hypothesis is not a vague, bold claim but at least roughly based on a solid theoretical framework (General Theory of Relativity, Quantum field theory) and on empirical constraints. Nevertheless, as already mentioned, Smolin's premises are quite speculative. And he himself admits that „what I am presenting ... is a frank speculation, if you will, a fantasy" [57, p. 7].

## 5.2 The View from Nowhere

Smolin claims that „the desire to be able to see the universe from the outside, as a disembodied observer ... is a remnant of the old physics and is inconsistent with relativity and quantum theory"; he propagates a view which describes „the universe as a coherent whole, in relationship only to itself, without need of anything outside itself to give it law, meaning or order" [57, p. 22]. He sees his hypothesis as „a new approach to the weaving of an objective view of the universe that, by its very construction, denies the possibility of its being read as the view of an observer outside the world" and postulates *„the principle of the logical exclusion of the possibility of an interpretation of cosmological theory in terms of an observer outside of the world"* [57, p. 251 and 334 resp.].
Smolin is certainly right that there is (for us) no God-like view from Nowhere down at the universe or Multiverse. However, the picture he himself gives us is also a view from „above". It is, in that respect, not different from other relativistic models. Speculations about the values of the physical parameters and the reproduction of other universes must describe these parameters and universes „from the outside" as long as we cannot avoid spatial descriptions at all. But this should be no problem – contrary to what Smolin might think. It is an abstraction or a metaphor which does not make his hypothesis right or wrong. It is still compatible with a relational view for which Smolin is arguing.

Smolin has, by the way, admitted [in personal communication] that he has not yet been able to formulate Cosmological Natural Selection in a language which fully satisfies his principle cited above. But perhaps a Quantum Theory of Gravity will provide such a language some day. There has been made in fact some progress in formulating an alternative to the Lorentzian space-time: the *theory of causal sets* [39-41]. Here, events in a discrete space-time, partially ordered by causal relations down to the Planck scale, can be defined prior to a space-time manifold which therefore appears as a classical limit only and might be described from an internal, finite-time point of view.

## 5.3    Predictability and Testability

According to Smolin, there are eight known variations in the values of fundamental parameters that lead to worlds with fewer black holes than our own, but there is no variation known with has clearly the opposite effect. This is already an interesting observation. Furthermore, Smolin's hypothesis has some predictive power, because there are physical effects and properties influencing black hole formation rates which are still not known (at least not precisely enough). Here, Smolin's hypothesis provides some constraints for these effects, if the number of black holes is almost maximized. These predictions can be tested in principle and partly even in the near future.

• The mass range of neutron stars provides the probably easiest test at the moment. If the mass of the $K^-$ meson (and hence the mass of the strange quark) is lower than a critical value, neutron stars could be made out of protons, neutrons and $K^-$ mesons inside, as Hans A. Bethe and Gerald E. Brown have suggested [5; 6], and not only out of neutrons as it is commonly thought (the electrons would turn to $K^-$ mesons and neutrinos instead of reacting with protons to neutrons and antineutrinos). If this is true, the upper mass limit of neutron stars is only about 1.5 solar masses. If the strange quark mass is higher than the critical value, neutron stars are made mostly out of neutrons as it is currently thought and could obtain masses up to, say, 2.5 solar masses. Because in this case the number of black holes in our universe would be at least ten times lower (more stars would collapse into neutron stars, not black holes), a high strange quark mass could be an indication that Smolin's hypothesis is wrong. (Note that this argument is independent of any Observational Selection effects associated with Anthropic reasoning, because the value of the strange quark mass may be varied within a large range before it produces a significant effect on chemistry.) Until yet it is not known if there are neutron stars with more than 1.5 solar masses. But if there are some, they could be easily detected (the mass of a neutron star can be measured very accurately if it is one of a pair of neutron stars in orbit around each other).

• On the other hand, if Smolin is right it should not be possible to further lower the upper mass limit of the strange quark below its actual value without somehow disrupting the process of star formation, because otherwise lower masses would imply an even higher black hole formation rate. However, such competing effects are still unknown and can only be determined by detailed calculations in nuclear physics.

• Other – much more crude – predictions are concerned with the variations of parameters like the mass of the electron, the strength of the weak interaction, the density of protons and neutrons, the duration of the presumed inflationary epoch of the early universe and so on.

• More difficult is this: There are some changes of a parameter which leads to different effects regarding the numbers of black holes produced (some effects increase the number of black holes while others decrease it). At the moment it is difficult to predict the overall outcome, given the complexity of the interrelations between these effects. Further studies will give a clearer picture. If it turns out that changes of these parameters decrease the number of black holes, Smolin's hypothesis will get into trouble. For instance an increase of the gravitational constant would decrease the mass of each star. Hence, more stars would be made out of the same given amount of mass. The result would be more black holes if everything else were the same. But it is not true that everything else stays the same in this case. Increasing the gravitational constant effects the formation of galaxies, the transfer of energy within stars, the evolution of stars, the amount of matter returned to the interstellar medium, the formation and lifetime of new stars and so on. And it is very difficult to predict the overall effect.

• There could be quite different universes with even more black holes, e.g. because of small primordial black holes due to compressions during an early phase transition at the beginning of the universe [52]. If there had been many more primordial black holes within the early universe which have already evaporated due to quantum-mechanical processes, they would still leave an imprint in the cosmic microwave background (and perhaps also in the gamma range). Given the accuracy expected for CMB observations from satellites starting within the next decade, there is a realistic possibility that observations will distinguish between different (but until yet not convincingly established) theoretical models of the early universe and its formation rate of primordial black holes.

• A faster star evolution or a different mechanism of early star formation could also refute Smolin's hypothesis. For example, in a world without stable nuclear bound states many more massive collapsed objects would become black holes than in our universe, where the collapse is delayed by stellar nucleosynthesis. On the other hand, catalytic effects of heavy elements are leading to a strong positive feedback in massive star formation (via cooling mechanisms of carbon monoxide) from the initial rate of massive star formation without heavy elements. The question is then, whether a combination of observational and theoretical progress could disentangle the opposite effects. If it turns out that the number of black holes formed is greater in a world without nuclear bound states than in our universe, then Smolin's hypothesis is in trouble.

• Cosmological Natural Selection requires only a fine-tuning of some parameters within a certain range in accordance to maximize black hole formation rates. There is no need for an even finer tuning as it might be necessary for a Theory of Everything. For example, the density parameter $\Omega$ (the quotient of the universe's mean mass density and its critical mass density), the cosmological constant $\Lambda$ [see 63; 65] and the neutrino masses [see 61; 71] are near a certain value which is criti-cal for a universe hospitable for earth-like life, but they need not have an exact value like Λ ≡ 0, Ω ≡ 1 etc.
• There are proposals of alternative theories, for example regarding a black hole's ability to create new universes, which are not consistent with Smolin's hypothesis [2].
In conclusion, Smolin's hypothesis makes some interesting and in principle testable predictions. They are at home in the reign of current cosmology and particle physics. Thus, Smolin's account is of scientific interest. It is not merely another myth of origin or a just-so story! However...

## 5.4 Smolin's central claim cannot be falsified

*Falsifiability* of a hypothesis depends on holding fixed the auxiliary assumptions needed to produce the targeted conclusion. In practice, one tries to show that the auxiliaries are themselves well confirmed or otherwise scientifically entrenched.
What should be falsifiable according to Smolin is his claim that *our* universe is nearly optimal for black hole formation. However, this is not a necessary consequence of his premises. A consequence is only that *most* universes are nearly optimal. To move from this statistical conclusion to the targeted conclusion about our universe, Smolin [57, p. 127] simply assumes that our universe is typical (see 2.4.3). This is an additional hypothesis as he admits. But this auxiliary assumption is neither confirmed nor otherwise scientifically entrenched! Thus, if changes in the values of our parameters did not lead to a lower rate of black hole formation – contrary to Smolin's prediction – we could always „save" Smolin's hypothesis by supposing that our universe is not typical, i.e. rejecting a Principle of Mediocrity.
Hence, there is (at least at the moment) no possibility to falsify Smolin's central claim that our universe is nearly optimal for black hole formation [cf. 75].
(One could introduce other *ad hoc hypotheses* as well in order to save Smolin's central idea. But this might undermine its falsifiability, too. For example, if there are parameters whose variation from their actual value leads to an increase of black hole formation, one could still claim that these parameters cannot be varied without also changing other parameters, leading e.g. to large side-effects in star formation, hence the originally varied parameters are no *independent* parameters contrary to the assumption; cf. 3.1.1.)
But note: if there is someday a Theory of Everything which uniquely determines the values of the physical parameters, the hypothesis of Cosmological Natural Selection would still be falsified at last.

## 5.5 Some problems and possibilities

Thus, Smolin's account is daring and speculative but not a pure and untenable speculation. He made interesting predictions about our universe's capability to form black holes which are or will be testable. It would be an important discovery if it turns out that our universe has indeed the property to produce an almost maximal number of black holes within a definite volume (or at least an almost maximal number of black holes of a certain mass within a definite volume). This would be a fact which certainly needs an explanation. However, even if Smolin is right and every prediction of his hypothesis could be verified, Cosmological Natural Selection wouldn't be confirmed – because there may be other explanations for the optimal black hole formation rate or none at all. It is still a question of the truth of Smolin's premises. And a maximal black hole formation rate of our universe cannot prove these premises.

What else may count as a prove for Smolin's premises then? Aren't they unverifiable and unfalsifiable in principle, not useful at all and hence just bad metaphysics in the tradition of medieval questions like asking for the highest possible number of angels on top of a needle? Not necessarily:

• Because there is a – nowadays still very vague – theoretical framework that might describe the origin of new universes someday; Smolin's premises might turn out as an implication of an otherwise well-established theory of quantum gravity in the future.

• Because one could still argue that such a hypothesis functions as an explanation of a set of observational data and that the other universes are the unobservables that are postulated to explain these observed data by showing them to be necessary rather than *a priori* improbable; to complain that the other universes are themselves unobservable would be to confuse the logical role of the *explanans* with that of the *explanandum*; it is the *explanandum* that needs to be observed, not the *explanans* [cf. 53, p. 345]. One might stipulate on this line of reasoning the existence of very weird things however. Thus, we need additional clues and arguments to provide a reasonable theoretical background.

• Because there might be observational effects of hypothetical connections between our universe and others (which might lead to a vanishing of the cosmological constant, for instance [8; 9]) or even traversable wormholes between them [70; 73]. But it would be indeed impossible to look at or into other universes, let alone to investigate the values of their fundamental parameters, if all effects and space-time tunnels are hidden behind event horizons of black holes.

• Finally, one could argue for a naturalized version of the principle of plentitude – that is to claim that everything exists what is not explicitly forbidden by the laws of physics [51, p. 108; cf. 29, p. 395; 30, p. 111]. In particle physics, for example, every process is treated as possible as long it is not excluded by conservation principles etc. However, this may only motivate the view that the assumption of other universes is not a priori meaningless. But it does not prove Smolin's specific hypothesis.

Nevertheless, Smolin's premises might turn out to be philosophically useful even if they are beyond science: They may help to carve out the space of possibilities for example as an antidote against an uncritical belief in idealism, teleology or a Theory of Everything. This is probably irrelevant for science (and should be not confused with rigorous scientific methodology, of course); but it may still inspire philosophical reasoning. It is also a fascinating example for reflections of the presuppositions and limitations of our intellectual and epistemological capabilities. And it is of cultural interest because it illustrates the power or influence of the new paradigm of self-organization which is compatible with an ontological reductionism without spooky metaphysical entities [66]. Last but not least,

Smolin's hypothesis has an heuristical value, because it offers a new way of looking at the parameters of particle physics and cosmology. Compared with all the other proposals so far offered to explain the fundamental physical parameters, Smolin's predictions (cf. 5.3) are the ones most vulnerable to falsification and hence well qualified for scientific testing. Therefore, they deserve further study.

# 6    Metaphysics

## 6.1    No refutation of the argument from design

Smolin's idea is an inspiring counterexample against the Argument from Design (also called the Teleological Argument) – the claim that the purposeful arrangement and architecture of the (structure, laws and parameters of the) universe points to a teleological force or creator, for instance God. As in biology, self-organized processes could lead to the emergence of complicated "teleonomical" (i.e. functionally or teleologically described) structures out of much simpler ones. Thus, it is not a necessary truth that our universe has a divine origin. Recent trials to revitalize the infamous Argument from Design as a proof for God (or some other teleological force) with the help of the fine-tuning of the universe [10; 17; 60; cf. also 32] are simply a *non sequitur*. The fact that we can *interpret* the universe as being designed, purposeful and meaningful does not imply that it actually *was* designed and has a purpose or meaning.

Of course, this sceptical line of reasoning is not new [cf. 14, ch. 7.3; 28, p. 130]. But Smolin has presented a scenario which is more than hand waving speculations. However, it is still possible to claim that the Multiverse or the quantum vacuum or the natural laws are of divine origin. It is impossible to refute a sufficiently sophisticated claim of the existence of hidden forces, and not only within science – but the burden of proof lies on the shoulder of its proponent anyway. Thus, Smolin's scenario cannot and need not replace philosophical criticisms of the Teleological Argument, but may support them.

## 6.2    No sufficient reason

Smolin's scenario cannot explain everything (nor does it want to, so this remark is not meant to criticize it but to stress an important limitation). Suppose, Smolin is right. Then we would know why the fundamental physical parameters are the way they actually are. They would be explained, or we would need no explanation anymore.
However, this merely shifts the crucial question further back, because
• we would still not know why the laws are the way they actually are – for instance, why quantum gravity implies the birth of universes out of black holes and why quantum gravity is true at all,

• we would still not know how the first universe (the „mother of all universes") came into being – or, if there is an infinite chain of universes, why there are universes at all.

Thus, Smolin's hypothesis does not give us a sufficient reason. Probably, a sufficient reason cannot be given at all [cf. 72, ch. 6] – which is not to deny that it may play an important heuristic role in the development of science. But Cosmological Natural Selection is not a First Cause or a principle proven by itself. It cannot help us to overcome contingency. It cannot answer why there is *anything* at all (nor can the speculations about cosmic engineers). And it cannot explain why the physical *laws* are the way they are. In the *ultimate* sense, the cosmos remains a mystery.

*Acknowledgements:* I am grateful to Peter C. Bosetti for the opportunity to contribute as well as Volker Müller and especially André Spiegel for valuable comments. It is also a pleasure to thank Lee Smolin for discussion.